\documentclass[fleqn,usenatbib]{mnras}

\usepackage{soul}

\usepackage[T1]{fontenc}

\DeclareRobustCommand{\VAN}[3]{#2}
\let\VANthebibliography\thebibliography
\def\thebibliography{\DeclareRobustCommand{\VAN}[3]{##3}\VANthebibliography}

\usepackage{graphicx}	
\usepackage{amsmath}	
\usepackage{amssymb}	

\newcommand{\orcid}[1]{\href{https://orcid.org/#1}{\includegraphics[scale=0.005]{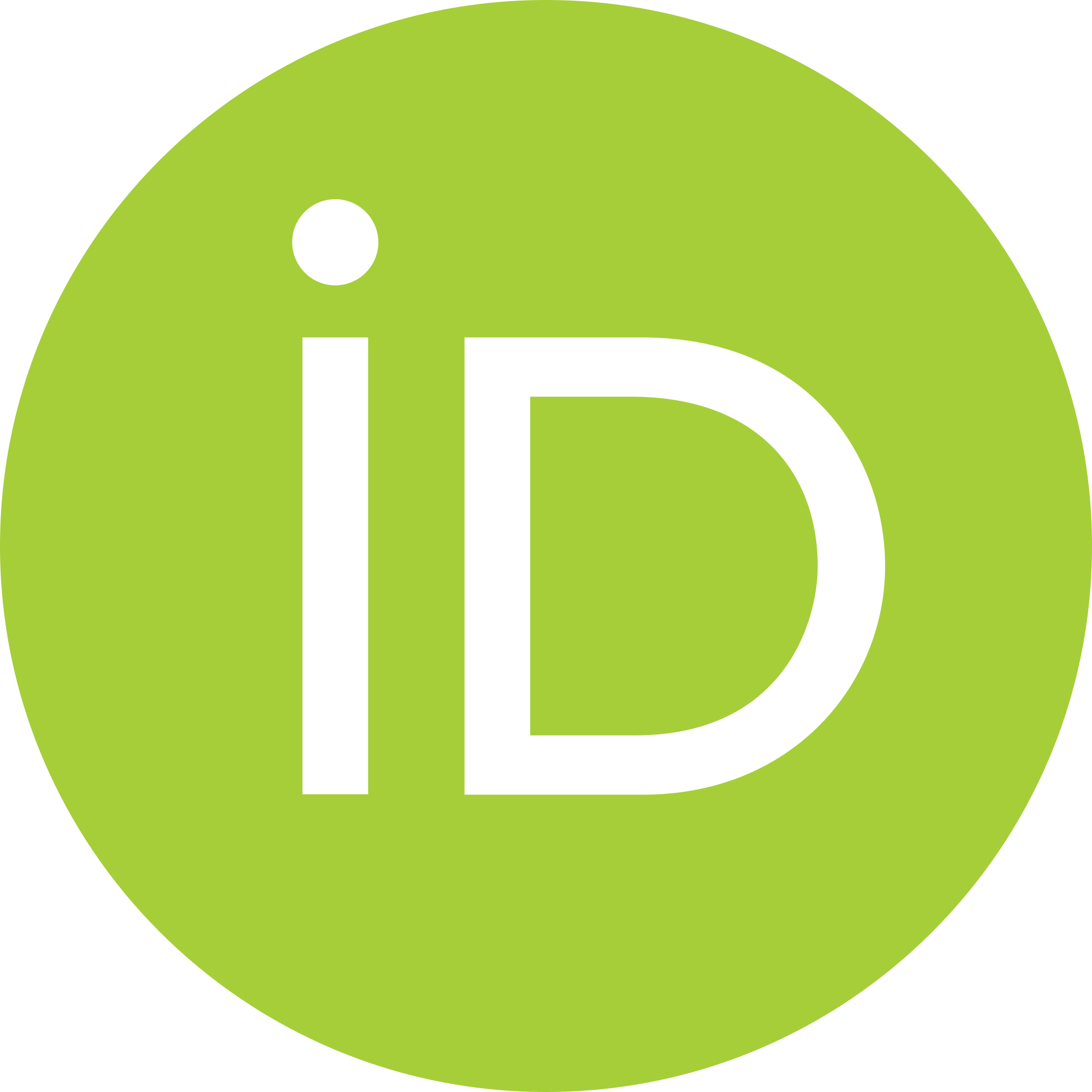}}}
\usepackage{newtxtext,newtxmath}

\title[Constraining disc mass]{Constraining protoplanetary disc mass using the GI wiggle}

\author[J. P. Terry et al.]{
J. P. Terry \orcid{0000-0002-8590-7271}$^{1, 2}$\thanks{E-mail: jpterry@uga.edu},
C. Hall \orcid{0000-0002-8138-0425}$^{1, 2}$, C. Longarini$^{3}$ \orcid{0000-0003-4663-0318},
G. Lodato \orcid{0000-0002-2357-7692}$^{3}$, 
C. Toci \orcid{0000-0002-6958-4986}$^{3}$, 
\hspace{2mm}B. Veronesi \orcid{0000-0002-2611-7931}$^{3,4}$, 
\newauthor T. Paneque-Carre{\~n}o \orcid{0000-0002-4044-8016}$^{5}$
and C. Pinte \orcid{0000-0001-5907-5179} $^{6,7}$ 
\\
$^{1}$Department of Physics and Astronomy, The University of Georgia, Athens, GA 30602, USA\\
$^{2}$Center for Simulational Physics, The University of Georgia, Athens, GA 30602, USA\\
$^{3}$Dipartimento di Fisica, Università degli Studi di Milano, via Celoria 16, 20133 Milano, Italy\\
$^{4}$ Univ Lyon, Univ Lyon1, Ens de Lyon, CNRS, Centre de Recherche Astrophysique de Lyon UMR5574, F-69230, Saint-Genis,-Laval, France\\
$^{5}$European Southern Observatory: Garching, Bayern, Germany\\
$^{6}$School of Physics and Astronomy, Monash University, Clayton Vic 3800, Australia\\
$^{7}$Univ. Grenoble Alpes, CNRS, IPAG, F-38000 Grenoble, France
}

\date{Accepted XXX. Received YYY; in original form ZZZ}

\pubyear{2021}

\begin{document}
\label{firstpage}
\pagerange{\pageref{firstpage}--\pageref{lastpage}}
\maketitle

\begin{abstract}
Exoplanets form in protoplanetary accretion discs. The total protoplanetary disc mass is the most fundamental parameter, since it sets the mass budget for planet formation.  Although observations with the Atacama Large Millimeter/Submillimeter array (ALMA) have dramatically increased our understanding of these discs, total protoplanetary disc mass remains difficult to measure. If a disc is sufficiently massive ($\gtrsim$ 10\% of the host star mass), it can excite gravitational instability (GI). Recently, it has been revealed that GI leaves kinematic imprints of its presence known as the ``GI Wiggle.'' In this work, we use numerical simulations to determine an approximately linear relationship between the amplitude of the wiggle and the host disc-to-star mass ratio, and show that measurements of the amplitude are possible with the spatial and spectral capabilities of ALMA. These measurements can therefore be used to constrain disc-to-star mass ratio.
\end{abstract}




\section{Introduction}

Arguably the most fundamental parameter of a protoplanetary disc is its total mass, since it determines the total amount of mass available for planet formation. However, direct measurements of disc mass remain elusive. At the cool temperatures protoplanetary discs exist at, molecular hydrogen \textemdash H$_2$, which constitutes between $\sim$90-99\% of the disc mass \textemdash cannot be observed. In order to circumvent this problem, disc masses are frequently estimated by converting continuum flux density at $\sim$mm wavelengths to a total dust mass. The total disc mass is then evaluated through the assumption of optically thin emission and a constant dust-to-gas ratio \citep{beckwith1990}. While this ratio is canonically assumed to match that of the ISM (1:100), disc measurements show a variety of deviations. First, the gas disc, measured in $^{12}$CO, typically extends somewhere between a factor of 2 and a factor of 4 out beyond the dust disc, as observed in the $\sim$mm continuum \citep{Panic_2009, Birnstiel_2014,Ansdell_2016, Pinte_2016, Facchini_2018, Toci_2021} due to inward radial drift of dust \citep{weidenschilling1977}. Secondly, if significant grain growth has taken place, this would shift more emission to longer wavelengths, and would require observations at longer wavelengths to recover the dust mass \citep{ilee2020}. It is therefore reasonable to assume that the dust:gas ratio is larger than the canonical ISM value in these regions. 

An alternative method is to measure line emission from molecules such as CO and its isotopologues \citep{miotello2014,Williams2014,miotello2016,Bergin_2017} and convert to total gas mass (or surface density \citealt{miotello2018}), through abundance ratios. Although generally thought to be more accurate than relying on dust-to-gas mass conversion, it is now understood that molecular abundances change in both space and time throughout a disc \citep{Ilee2017, quenard2018, zhang2019}, and so this method is likely subject to similar uncertainties driven by local changes to abundance ratios.

Recent near-infrared and sub-millimetre observations of protoplanetary discs (see, e.g., \citealt{Benisty2015, Perez2016, Andrews2018, Huang2018, Benisty2021})  have revealed prominent spiral structure in multiple discs, which may, in some cases, be due to gravitational instability (GI)~\citep{Dong2015,Hall2016,Meru2017, Veronesi2019, hall20,  cadmanhall2020, Chen21}. While it has been demonstrated that GI can be responsible for spiral morphology of some observed discs, it requires the disc-to-star mass ratio, $q$, be $\gtrsim 0.1$ for the spirals to be observable \citep{Cossins2010, Dipierro2014, Dong2015, Kratter16, hall2018, Hall2019}. The presence of GI, by definition of its existence, therefore places constraints on the mass of a disc.

An interesting possibility with GI discs is fragmentation. Essentially, if a disc is sufficiently massive, and able to cool sufficiently quickly,  then a region of that disc may fragment to form gravitationally bound objects \citep{Gammie2001,rice2003a,Rice05}. Fragmentation has been proposed as a complementary planet formation pathway to the standard core accretion paradigm~\citep{Boss1997, Boss1998, Nayakshin2010}, that could offer a plausible explanation for massive objects on wide orbits such as those around HR 8799 \citep{hr8799}, the potential companion object in TW Hya \citep{tsukagoshi2019,nayakshin2020}, and the massive objects potentially forming in AB Aurigae \citep{cadmanhallrice2021}.

However, in general, if such objects regularly form they are likely to evade detection with instruments such as ALMA \citep{humphries2021}. Analytical calculations \citep{rafikov2005}, population synthesis models \citep{forganrice2013, forgan2018b} and hydrodynamical simulations \citep{hall2017} indicate that GI planet formation is most likely to result in objects $\geq 5$M$_{J}$ at distances $\geq$ 50 au from their host star \citep{rice2015}. Furthermore, it has recently been shown that less massive discs are more stable to GI \citep{Haworth2020}, which may cause fragmentation to occur preferentially around the most massive objects \citep{ilee2018, Cadman2020}.

In either case, measuring the disc mass is crucial for determining the dynamical fate of the system - i.e., fragmentation \citep{Gammie2001,rice2003a,rice2003b}, quasi-steady GI \citep{Lodato04}, episodic GI \citep{Lodato05} (see \citet{Kratter16} for a review of these topics) - and the total mass budget available for planet formation. Analysis of the disc rotation curve can give insight into this parameter. Recently, dynamical measurements of disc mass have been obtained through observing deviation from Keplerian behaviour ($v_\phi = \left[ \rm{GM_{*}/r} \right]^{1/2}$) in the rotation curve of Elias 2-27 \citep{veronesi2021}.

Morphologically, it has been observed that there is an approximate relationship between the disc-to-star mass ratio, $q$, and the number of spiral arms, $m$, such that $m\sim 1/q$ \citep{cossins2009,Dong2015,Hall2019}. However, simulated observations have shown that for an instrument such as ALMA, the correct number of spiral arms will not always be recovered from the observation \citep{Dipierro2014, Hall2019}, which depends not only on the resolution, but also on the arm to inter-arm contrast ratio \citep{Hall2016}. Therefore, this $m\sim1/q$ method cannot be relied upon to infer total disc mass from observations.

Recently, \citet{hall20} presented a prediction for the kinematic signature of GI known as the ``GI Wiggle''. In individual line emission velocity channels, it is a distinctive ``zig-zag'' feature. When Keplerian rotation is subtracted from the intensity-weighted velocity, the wiggle appears as ``interlocking fingers''~\citep{hall20}. It is caused by sustained velocity perturbations throughout the disc, above and below the average azimuthal velocity inside and outside the disc spiral arms. Recent observations have determined that there is evidence of this feature in the Elias 2-27 system \citep{paneque2021}. 

\begin{figure}
    \centering
    \includegraphics[width=0.99\columnwidth]{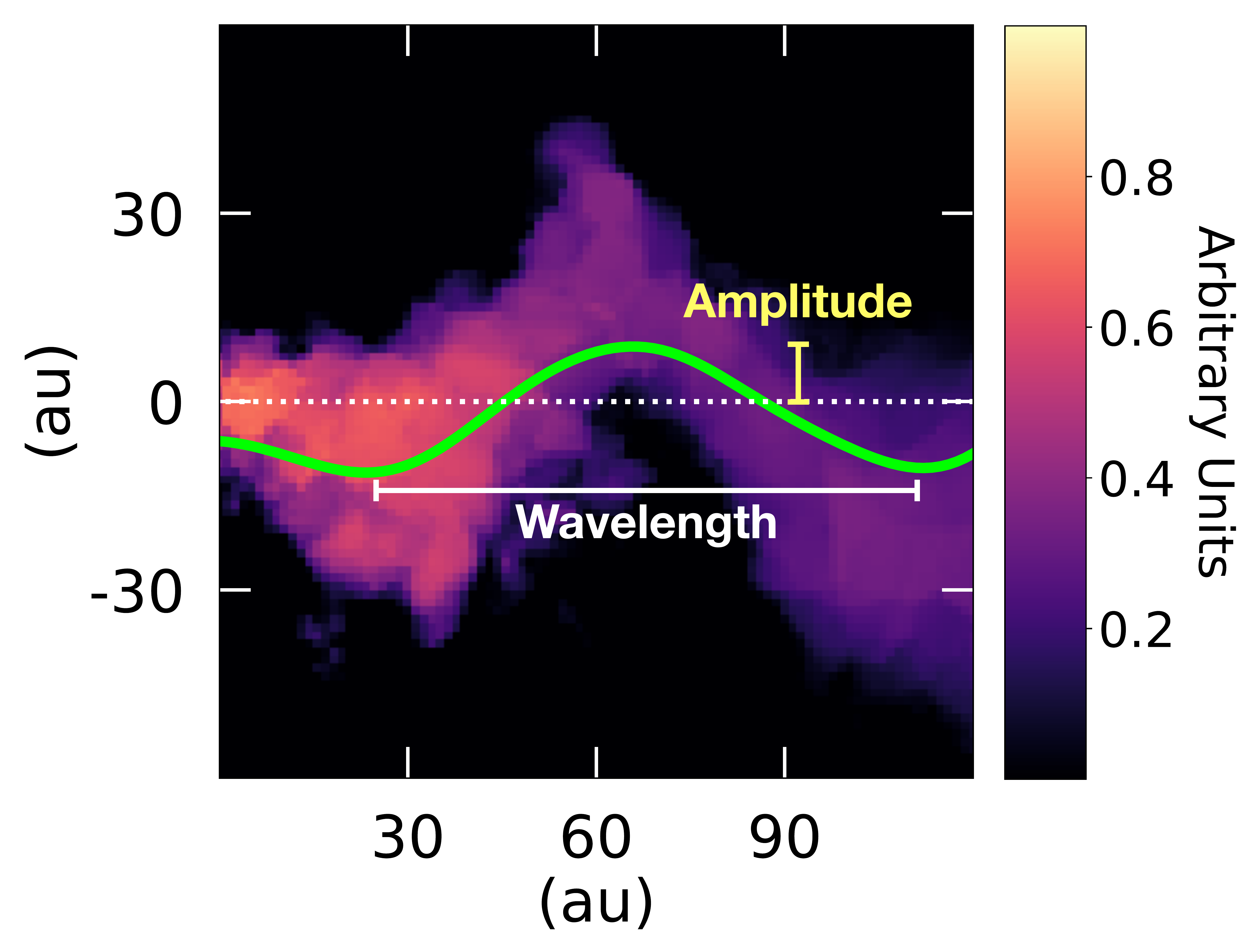}
    \caption{Illustration of the dominant amplitude (A$_{\rm{wiggle}}$) and wavelength ($\lambda_{\rm{wiggle}}$) of the GI wiggle overlaid on the synthetic line emission observation from which the signal was extracted. The systemic velocity channel is used ($v_{\mathrm{obs}} = v_{\mathrm{systemic}}$). For clarity, the displayed signal has been rotated to 90$^{\circ}$ relative to Figure~\ref{fig:images} and now travels radially outwards from the star.}
    \label{fig:schematic}
\end{figure}

Our aim in this work is to present, through numerical simulations, the relationship between the disc-to-star mass ratio, $q$, and the ``strength'' or ``amplitude" of the GI Wiggle. Kinematic analysis is a promising avenue for extracting system properties, as deviations from Keplerian velocities within the disc can be linked to embedded objects (e.g. protoplanets) or physical processes (e.g. GI) that influence the system evolution~\citep{Perez_2015, Teague2018, Pinte2018, Pinte2020, hall20, Armitage2020, paneque2021, Bollati2021, Wolfer2021}. It has been shown by~\citet{hall20} that, in contrast to local velocity fluctuations caused by the spiral wake of a protoplanet, (see, e.g., \citealt{Pinte2018}) GI-driven spirals result in global velocity perturbations (``GI wiggles") with a high degree of rotational symmetry, and a more uniform quality than (for example), perturbations caused by a companion of larger planet mass \citep{Perez_2018},  or a vortex induced by a Rossby wave instability \citep{huangp2018}.

Another process, the vertical shear instability (VSI) \citep{Nelson2013}, may, like GI, be able to induce perturbations with a high degree of rotational symmetry \citep{barraza2021}. However, these perturbations are an order of magnitude smaller than those induced by GI ($\sim$0.06 km/s compared to $\sim$0.4 \citealt{barraza2021,hall20}), so it should be possible to differentiate between them. Finally, a new and promising rotation curve technique demonstrated by \citet{veronesi2021} would\textemdash when coupled with the presence of the GI-Wiggle \textemdash provide very strong evidence of GI over VSI.

\begin{figure*}
    \centering
    \includegraphics[width=0.9\linewidth]{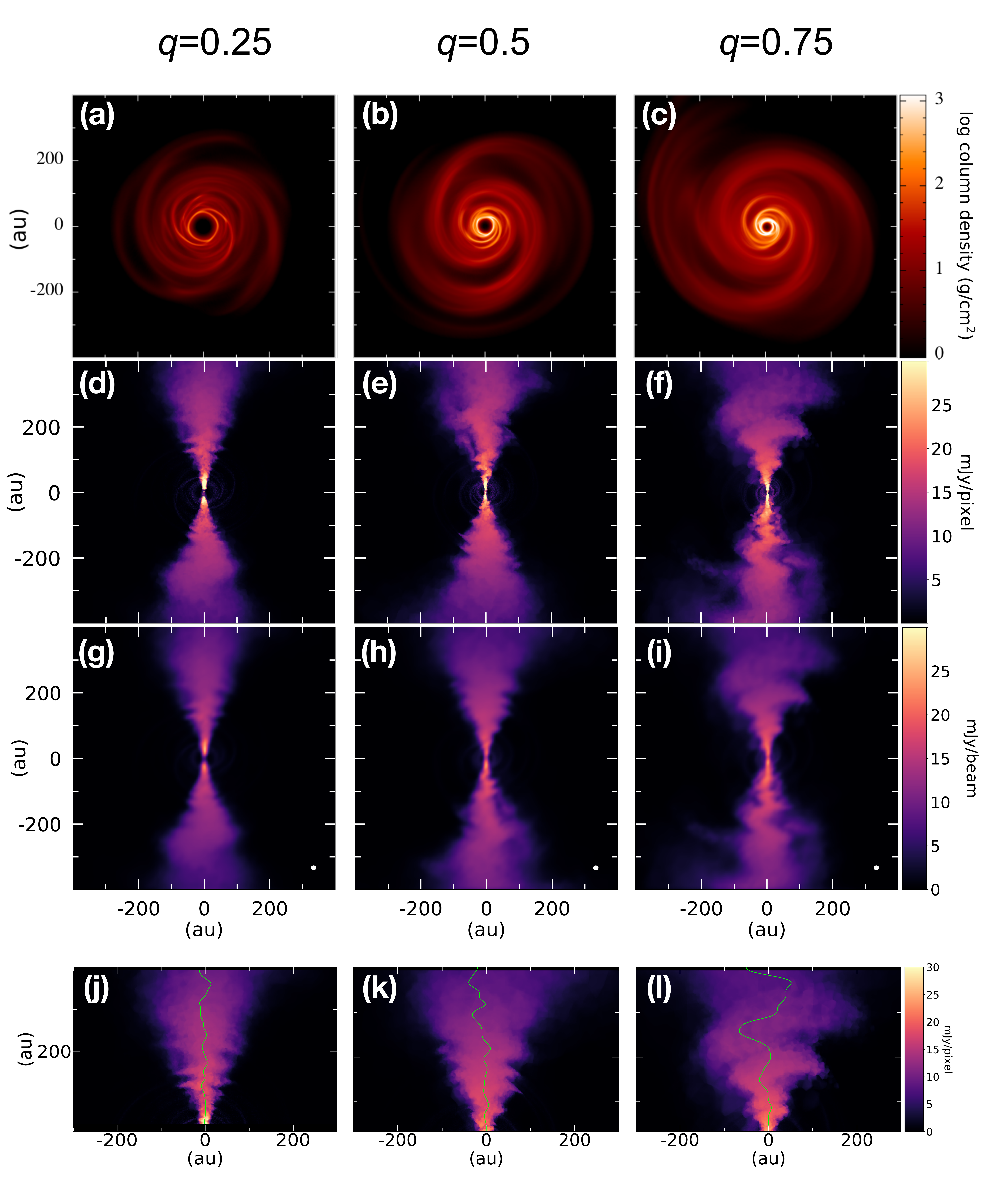}
    \caption{Hydrodynamical models of self-gravitating discs and associated GI wiggle. From left to right, $q$=0.25, $q$=0.5, $q$=0.75. Top Row: SPH log column density [g cm$^{-2}$].
             Second Row: $^{13}$CO line emission in mJy/pixel from systemic ($\Delta v \equiv v_{\mathrm{obs}} - v_{\mathrm{systemic}} = 0$ km/s) velocity channels from radiative transfer calculations on raw simulation results. Each pixel is 1.33 au by 1.33 au. 
             Third Row: $^{13}$CO line emission in mJy/beam from $\Delta v=0$ km/s velocity channels after spatial and spectral convolution with beam size of 7.7 au by 5.6 au, indicated in the lower right corner.
             Bottom Row: Extracted signals overlaid on the $\Delta v = 0$ km/s velocity channel. Note that these are rotated relative to the signal in Figure~\ref{fig:schematic}.}
    \label{fig:images}
\end{figure*}

We describe the wiggle in terms of the parameters wavelength and amplitude. Figure~\ref{fig:schematic} shows these for a representative systemic velocity channel ($v_{\mathrm{obs}} = v_{\mathrm{systemic}}$). The perturbations corresponding to the amplitude are in the azimuthal direction, and the wavelength is measured radially from the central star. Figure~\ref{fig:images} further illustrates this. We use these parameters to obtain the relationship between the wiggle and the disc-to-star mass ratio. 
 
The structure of the paper is as follows: Section \ref{sec:methods} describes the methods for the simulations and our analysis; Section \ref{sec:results} presents the results; Section \ref{sec:discussion} discusses the results and the limitations of the study; In Section \ref{sec:conclusion}, we offer our conclusions.

\section{Methods}
\label{sec:methods}

\subsection{Hydrodynamical simulations}
\label{subsec:sims}

We performed simulations of three-dimensional self-gravitating protostellar discs composed of dust and gas using the \texttt{Phantom} Smoothed Particle Hydrodynamics (SPH) code \citep{phantom18}. Dust was modelled self-consistently with the gas, using the ``one-fluid'' method \citep{laibeprice2014a, laibeprice2014b, laibeprice2014c, hutchison2018}, which in practice relies on the terminal velocity approximation \citep{youdin2005}. The back reaction of the dust onto the gas is included, and we use the flux-limited prescription of \citet{ballabio2018}. We allowed $\alpha_{\rm{SPH}}$ to vary between 0.1 and 1.0 and set $\beta_{\rm{SPH}} = 2.0$. The $\alpha_{\rm{SPH}}$ values correspond to Shakura-Sunyaev viscosities, $\alpha_{\rm{SS}}$, of $\alpha_{\rm{SS}}\approx 0.0025-0.025$.  We assumed an initial dust-to-gas mass ratio of $\epsilon=0.01$, and followed the spatial evolution of dust divided into 5 different size bins from 1 micron to 4 mm. The initial dust distribution was set to be perfectly mixed with gas. Even though we focus in this work on the synthetic CO emission, we include dust since it allows us to accurately determine the temperature structure of the disc in the radiative transfer calculation (see Section \ref{subsec:mcfost}).

Seven simulations were performed in total, with disc-to-star mass ratios of $q=0.15,0.2,0.25,0.4,0.5,0.6$ and $0.75$. As shown in Table~\ref{tab:r_outer}, the number of SPH particles was varied to maintain the relationship between scale height, $H$, and smoothing length, $h$, such that $h/H<0.25$ for the majority of the disc (shown in Appendix Figure \ref{fig:h_H}), satisfying resolution requirements as outlined in \citet{Nelson2006, lodatoclarke2011}. This has the additional effect of roughly maintaining the artificial viscosity as disc mass increases, a measure of which is seen in Appendix Figure~\ref{fig:cube_root_fit}. See the Appendix for a more thorough discussion on the selection of the number of SPH particles for each $q$. The main simulation parameters are shown in Table \ref{tab:r_outer}.

\begin{table}
 \centering
 
 \begin{tabular}{lccc}
  \hline
  $q$ & $R_{\rm{outer}}$ (au) & $R_{\rm{res}}$ (au) & N\\
  \hline
    0.15 & 153 & 32 & 500,000\\
    0.2 & 174 & 31 & 500,000\\
    0.25 & 186 & 28 & 750,000\\
    0.4 & 213 & 24 & 1,000,000\\
    0.5 & 216 & 23 & 1,250,000\\
    0.6 & 234 & 21 & 1,250,000\\
    0.75 & 237 & 13 & 1,500,000\\
  \hline
 \end{tabular}
\caption{Disc $R_{\rm{outer}}$, $R_{\rm{res}}$, and N (number of particles) for all $q$.
\label{tab:r_outer}}

\end{table}

The inner and outer radii of the disc were set to 10 and 300 au, respectively. 
The central star is represented as a sink particle \citep{bateetal1995} of mass $M_*=0.6$ M$_{\odot}$ with an accretion radius of 1 au. The gas surface density profile is $\Sigma \propto R^{-1}$ and the sound speed profile is $c_{\rm{s}}\propto R^{-0.25}$, consistent with both observational results~\citep{Perez2016, Huang2018, paneque2021, veronesi2021} and previous simulations of self-gravitating discs aimed at reproducing observations \citep{Meru2017, tomida2017, hall2018, Forgan2018}.

\par
The disc was set such that it was initially stable to GI, with the Toomre parameter \citep{Toomre1964}
\begin{equation}
    Q=\frac{c_s\kappa}{\pi\mathrm{G}\Sigma},
\end{equation}
set to $Q\gtrsim 2$. Here,  $\kappa$ is the epicyclic frequency, G is the gravitational constant, $c_{s}$ is the sound speed, and $\Sigma$ is the surface density. In keplerian rotation, $\kappa$ is simply $\Omega$. In the regime where $Q\lesssim 1$, the disc becomes unstable. If cooling processes balance heating processes, GI will efficiently transport angular momentum throughout the disc on long timescales \citep{Gammie2001, cossins2009, Forgan2011}. Alternatively, if the cooling rate is high, the disc may fragment to form gravitationally bound objects at the local Jeans mass \citep{Gammie2001}. For a recent review, see \citet{Kratter16}. The disc was allowed to cool through the simple $\beta$-cooling prescription \citep{Gammie2001}, where the cooling timescale is given by $t_\mathrm{c} = \beta /\Omega$. In all simulations, $\beta=15$, as in ~\citet{hall20}. The resulting surface density is shown in the top row of Figure~\ref{fig:images} for examples of the resulting column density.

\subsection{Thermal disc structure and $^{13}$CO channel map}
\label{subsec:mcfost}

The thermal disc structure and $^{13}$CO $J=3\rightarrow 2$ channel map was computed using MCFOST, a Monte Carlo radiative transfer code~\citep{Pinte2006, Pinte2009}. $^{13}$CO was chosen over $^{12}$CO as it is less likely to be affected by foreground contamination. This is particularly important if the disc is still at least partially embedded, as is likely the case for young self-gravitating discs.

We assumed that the gas and dust temperatures were equal and used $10^8$ photon packets to calculate dust temperature. As was done in the SPH simulations, the dust-to-gas ratio was 1:100. The density structure was obtained through direct Voronoi tesselation of the SPH particles, where each SPH particle corresponds to one MCFOST cell. Dust was assumed to be a mixture of silicates and carbon~\citep{Draine1984}, and the optical properties were calculated using Mie theory~\citep{Andrews_2009}. The dust grain sizes vary between 0.03 $\mu$m and 4 mm, following a logarithmic distribution split into 100 bins. At any location in the model, the density of the dust of grain size $a_i$ was obtained by interpolation from the SPH dust sizes. Any dust smaller than half the smallest SPH dust grain size (0.5 $\mu$m) was assumed to be perfectly coupled to the gas. The maximum size was set to be equal to the largest dust grain size in the SPH simulation (4 mm). Finally, the size distribution of the dust was normalised by integrating over all grain sizes assuming that the number density of dust grains, $n(a)$, was given by d$n(a)\propto a^{-3.5}$ d$a$.

The $^{13}$CO abundance relative to H$_{2}$ was set to 7$\times 10^{-7}$ as in~\citet{hall20}. We set the stellar parameters to match those of the Elias 2-27 system as described in in~\citet{Andrews_2009}: T$_{*}$ = 3850 K, R$_{*}$ = 2.3 R$_{\odot}$, and M$_{*}$ = 0.6 $M_{\odot}$. The flux was synthetically simulated at a distance of 140 pc and an inclination of 30$^{\circ}$. Two sets of channel maps were produced. The "raw" data, with no spectral or spatial convolution (examples in second row in Figure~\ref{fig:images}), and a set where the channels are spectrally and spatially convolved (examples in third row in Figure~\ref{fig:images}). The convolved channel maps were generated using Hanning-smoothing at a spectral resolution of 0.03 km/s. A turbulent velocity of 0.05 km/s was assumed. The maps were then spatially convolved with a Gaussian beam of size $0.11\arcsec \times 0.07\arcsec$ and a position angle of $-38\deg$. This matches the expected spatial and spectral resolution necessary to kinematically detect a planet \citep{Pinte2019}, although it may be possible to detect such a signature at lower spatial and/or spectral resolution. The GI Wiggle is a somewhat larger feature than a planet-induced kink, and should therefore be readily detectable at this spatial and spectral resolution.

\subsection{Velocity perturbations}

\begin{figure}
    \centering
    \includegraphics[width=0.95\linewidth]{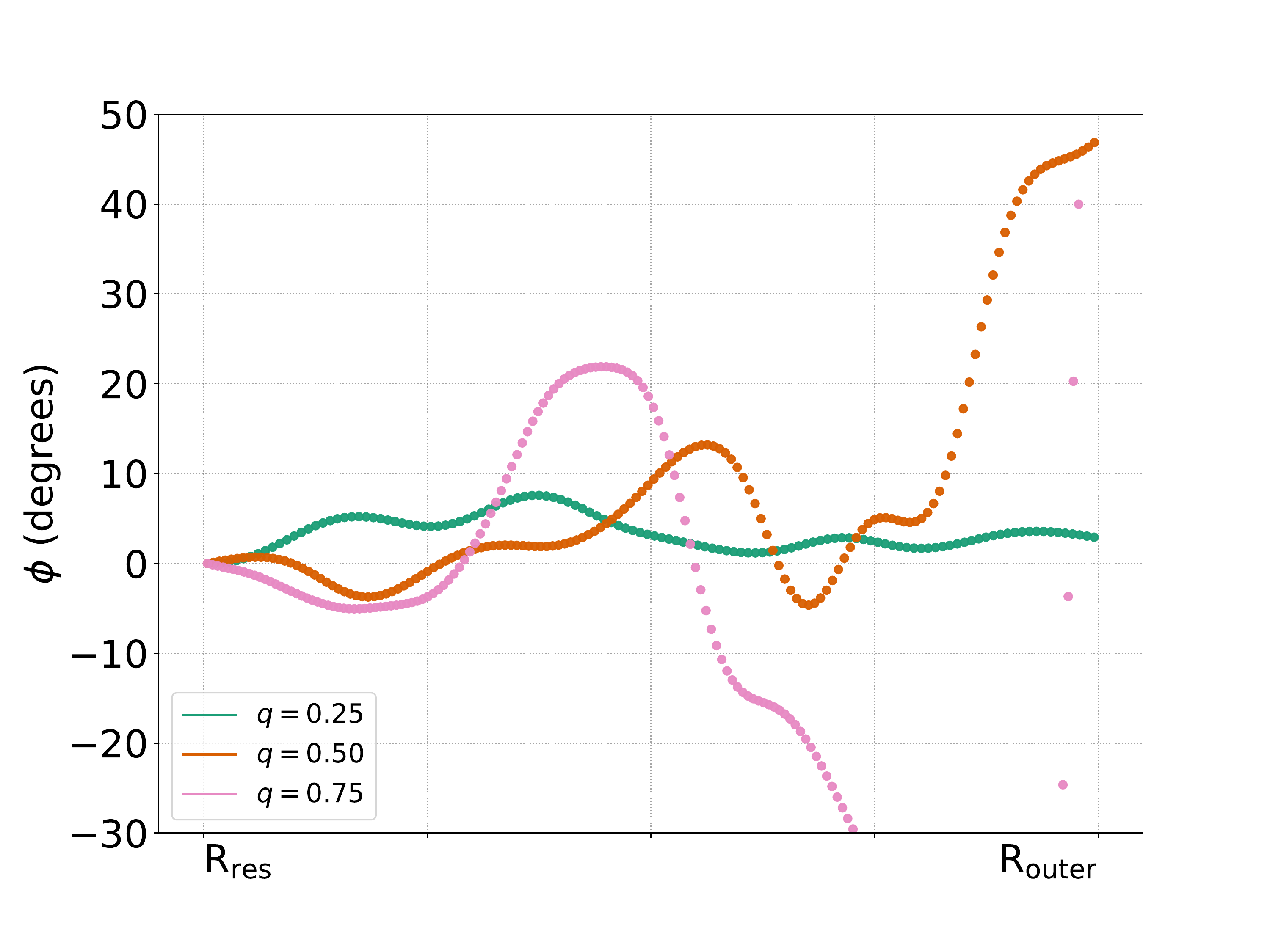}
    \caption{Smoothed GI wiggle signals from the $\Delta v=0$ km/s channel in polar coordinates for selected mass ratios. Signals trace the spine of $^{13}$CO emission and are extracted using Equation~\ref{eq:signal}. All signals are between each disc's $R_{\rm{res}}$ and $R_{\rm{outer}}$ to ensure the disc is well-resolved at that distance.}  
    \label{fig:signals}
\end{figure}

An observed velocity field is be described by a decomposition into the azimuthal, radial, vertical, and systemic motion using 
\begin{equation}
v_{\mathrm{obs}} = v_{\phi}\mathrm{sin}(i)\mathrm{cos}(\phi) + v_{\mathrm{r}}\mathrm{sin}(i)\mathrm{sin}(\phi) + v_{\mathrm{z}}\mathrm{cos}(i) +v_{\mathrm{systemic}}.
\label{v_obs}
\end{equation}

In the case of a purely Keplerian disc, $v_r=v_z=0$, $v_\phi = \sqrt{GM_{*}/r}$, and the observed velocity field is the well known ``butterfly pattern." When a self-gravitating spiral is present, the velocity field is perturbed. This perturbation results in a GI wiggle, which is a deviation from Keplerian rotation such that $v_r$ and $v_z$ become non-zero, and $v_{\phi} \neq  \left[ \rm{GM_{*}/r} \right]^{1/2}$.

When observed in line emission, this is a wave-like perturbation and therefore characterised by two properties: wavelength and amplitude, as illustrated in Figure~\ref{fig:schematic}. To extract wavelength and amplitude, we first define the signal, given by:
\begin{equation}
y_{n} = \frac{1}{I_{n} + \sum_{i=1}^{5} (I_{i} + I_{-i})}\left[ I_{n}y + \sum_{i=1}^{5} \bigg(I_{i}(y+i) + I_{-i}(y-i)\bigg) \right],
\label{eq:signal}
\end{equation}
where $y_{n}$ is the signal (in au). For the $n^{\rm{th}}$ row in a given line emission velocity channel (e.g. a row in Figure~\ref{fig:images} d, e or f), we locate the pixel (column in $x_{n}$), $y$, with the strongest emission, $I_{n}$. To minimise noise, we weight $y_{n}$ by the intensities of the 5 pixels on either side. A given pixel that is $i$ columns away from $y$ in $x_{n}$ has intensity $I_{i}$ and is located at $y+i$. For examples of extracted signals, Figure~\ref{fig:images} shows extracted signals overlaid on the line emission, and Figure~\ref{fig:signals} shows signals in polar coordinates. Note that the signals in all figures have been convolved with a Gaussian (i.e. Gaussian smoothing) for clarity.

\begin{figure*}
    \centering
    \includegraphics[width=0.95\linewidth]{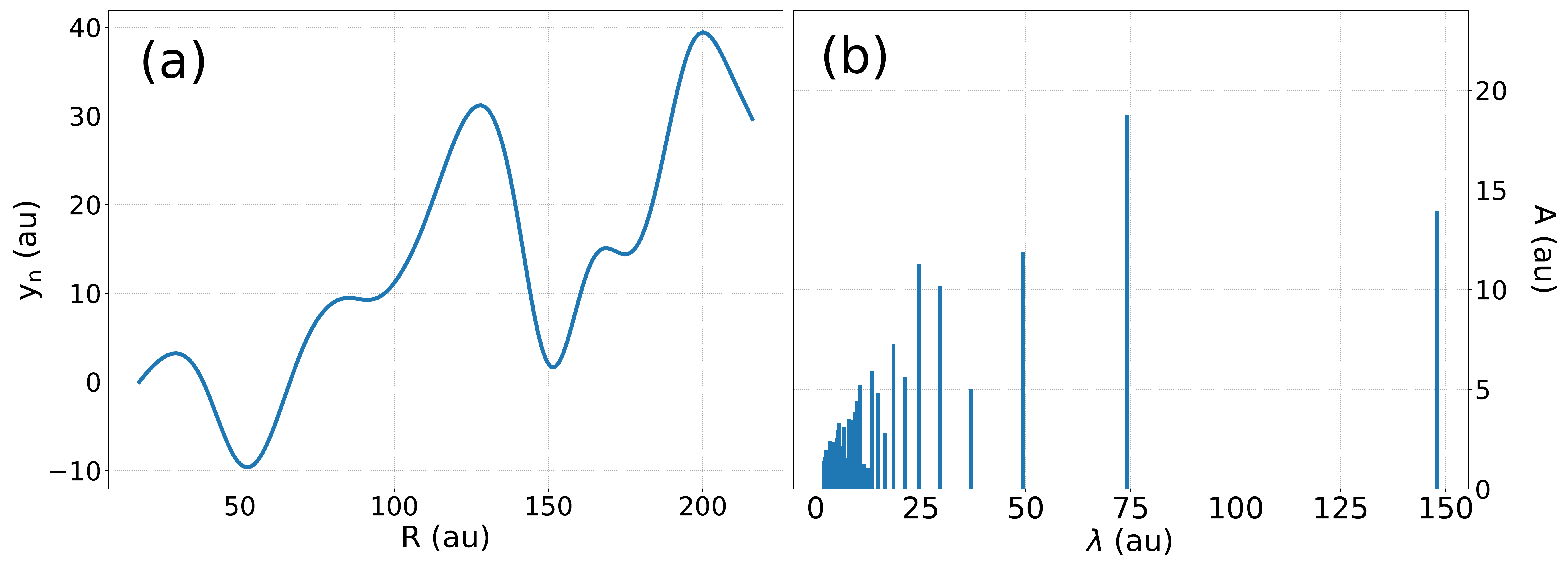}
    \caption{(a) Smoothed, intensity-weighted extracted signals from the $\Delta v = 0$ km/s channel for $q=0.5$. (b) Amplitude vs wavelength after Fourier decomposition. For this channel, $\lambda=74$ au is the dominant wavelength with an amplitude of A $=18.7$ au.}
    \label{fig:signal_ft}
\end{figure*}

Since the perturbation is approximately sinusoidal, it is well-described using a Fourier series. The extracted wiggle signal is decomposed using a Fourier transform into its power spectrum, $|A(\lambda)|^{2}$, for a given wavelength, $\lambda$. $|A(\lambda)|$ can then be interpreted as the amplitude of the component of the signal with that wavelength. Figure~\ref{fig:signal_ft} shows this decomposition. We define the wave amplitude, A$_{\rm{wiggle}}$, as max($|A(\lambda)|$) and $\lambda_{\rm{wiggle}}$ as the corresponding wavelength. 

\par
We only consider signals at radial distances, $R$, in the range $R_{\rm{res}} \leq R \leq R_{\rm{outer}}$, where $R_{\rm{outer}}$ is the outer radius containing 95\% of the disc mass and $R_{\rm{res}}$ is the shortest radial distance from the central sink such that $h/H < 0.25$ for all $R_{\rm{res}}\leq R \leq R_{\rm{outer}}$ (see Appendix Figure \ref{fig:h_H}). These values, along with the number of particles used for each simulation, can be found in Table ~\ref{tab:r_outer}. This method ensures that all extracted signals come from regions of the disc that are properly resolved. We apply this method over a range of velocity channels ($|\Delta v| = |v_{\mathrm{obs}} - v_{\mathrm{systemic}}| < 1$ km/s) that covers the the majority of the disc. We do not consider higher velocity channels due to the difficult of extracting the signal. The average dominant wavelength and amplitude of the velocity perturbation is determined for a given $q$. Results are also normalised by $R_{\rm{outer}}$ to present a general result that can be extrapolated to other systems. 

\par

Determining $R_{\rm{outer}}$ consistently from real observations is, of course, challenging. One could use either the size of the continuum emission or that of line emission. The dust size measured from continuum emission is typically a factor 2-3 smaller than the gas size measured from $^{12}$CO emission lines~\citep{Ansdell_2018, Sanchis_2021, Tazzari_2021}, due to a combination of radial drift and possibly substructure formation~\citep{rosotti2019b,Toci_2021}. CO-based measurements are generally more reliable than dust size measurements, but one should be careful that, due to optical thickness effects, the radius enclosing 95\% of the $^{12}$CO flux contains much more than 95\% of the mass. Using the 68\% CO flux radius may trace the bulk of the gas mass more accurately~\citep{Trapman_2020, Toci_2021}.

As described in Section \ref{subsec:mcfost}, two sets of channel maps are produced. The first is the ``raw'' data - i.e. the exact GI wiggle signal (second row in Figure~\ref{fig:images}). We first perform the analysis on the raw data (obtaining wavelength and amplitude), then repeat the analysis on the spatially and spectrally convolved channel maps (third row in Figure~\ref{fig:images}), to ensure that it would be possible to extract these properties from an observed system with current spatial and spectral resolution.

\section{Results}
\label{sec:results}

\begin{figure*}
 \includegraphics[width=\linewidth]{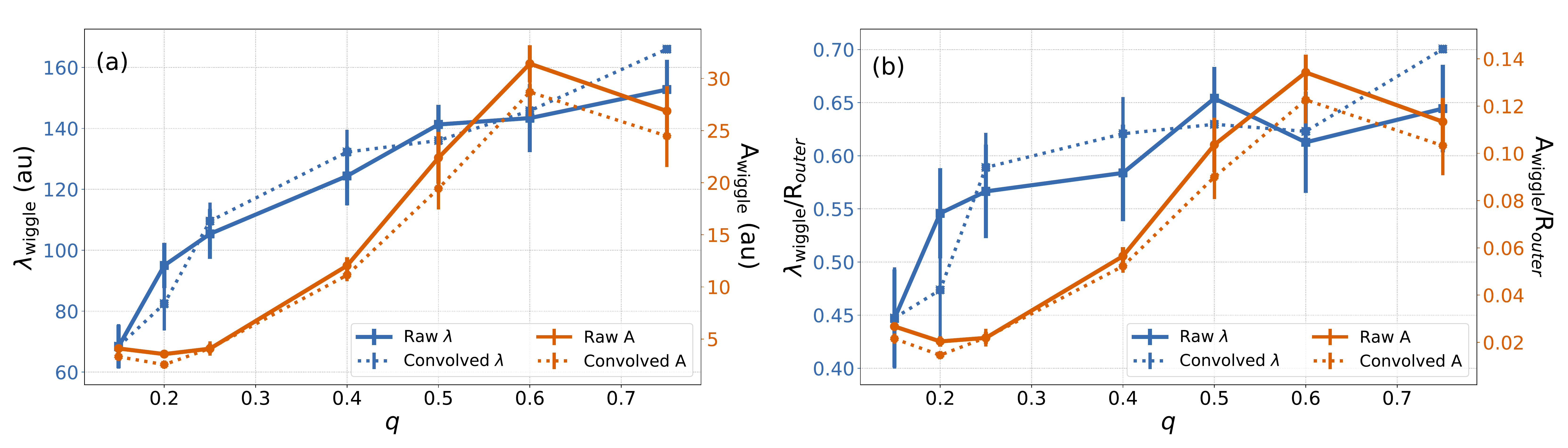}
 \caption{Amplitude (orange circles) and wavelength (blue squares) from the raw simulation data (solid line) and results that have been spatially and spectrally convolved (dotted line) plotted with standard errors, which are mostly smaller than the points. (a) Amplitude and wavelength in terms of au; (b) Amplitude and wavelength normalized by $R_{\rm{outer}}$. Standard errors, $\sigma/\sqrt{n}$, for a given distribution are plotted.}
 \label{fig:all_results}
\end{figure*}

\begin{table*}
 \centering
 
 \begin{tabular}{lccc}
  \hline
  Value & Fit Slope: m & Fit Intercept: b & R$^{2}$ \\
    & (au) & (au) & \\
  \hline
  Raw Amplitude & $50 \pm 8$ & $-5.4 \pm 3.8$ & 0.88 \\
  
  Raw Wavelength & $129 \pm 20$ & $66.0 \pm 9.3$ & 0.89 \\
  
  Convolved Amplitude & $48 \pm 9$ & $-6.6 \pm 4.5$ & 0.87 \\
  
  Convolved Wavelength & $133 \pm 20$ & $68.7 \pm 9.7$ & 0.92 \\

  \hline
  \end{tabular}
    \caption{Fit parameters and metrics from performing a linear $y=mx+b$ fit on A$_{\rm{wiggle}}$ and $\lambda_{\rm{wiggle}}$ as functions of $q$.} 
\label{tab:fit_results}
\end{table*}

\begin{table*}
 \centering
 \begin{tabular}{lccc}
  \hline
  Value & Normalized Fit Slope, m & Normalized Fit Intercept, b & Normalized R$^{2}$\\
  \hline
  Raw Amplitude & $0.200 \pm 0.037$ & $-0.013 \pm 0.017$ & 0.85 \\
  
  Raw Wavelength & $0.27 \pm 0.078$ & $0.471 \pm 0.035$ & 0.70 \\
  
  Convolved Amplitude & $0.196 \pm 0.040$ & $-0.021 \pm 0.020$ & 0.85\\
  
  Convolved Wavelength & $0.308 \pm 0.088$ & $0.467 \pm 0.043$ & 0.75\\

  \hline
  \end{tabular}
    \caption{Fit parameters and metrics from performing a linear $y=mx+b$ fit on A$_{\rm{wiggle}}$/$R_{\rm{outer}}$ and $\lambda_{\rm{wiggle}}$/$R_{\rm{outer}}$ as functions of $q$.} 
\label{tab:norm_fit_results}
\end{table*}

Our results show that there is a strong positive correlation between $q$ and GI wiggle amplitude, shown in Figure~\ref{fig:all_results}. Plot (a) shows the values in au, and plot (b) shows the values normalised by $R_{\mathrm{outer}}$ so that they can be generalised to discs of any size. Blue lines with squares are wiggle wavelengths, and orange lines with circles are wiggle amplitudes. Solid lines are raw data, while dotted lines are spatially and spectrally convolved data.  We quantify our results by performing an R$^{2}$ linear regression analyses, shown in Table~\ref{tab:fit_results} and Table~\ref{tab:norm_fit_results}.

\par
When we normalize our results with respect to $R_\mathrm{outer}$, shown in Figure~\ref{fig:all_results}(b), the positive correlation between amplitude and $q$ remains, although the R$^{2}$ values in Tables~\ref{tab:fit_results} and~\ref{tab:norm_fit_results} show that it is less robust. However, the wavelength as a fraction of $R_{\rm{outer}}$ substantially weakens the correlation with $q$. While the relevant R$^{2}$ values in Table~\ref{tab:norm_fit_results} are still large enough for a linear approximation to be acceptable, they are significantly lower than the equivalent values in Table~\ref{tab:fit_results}.

\section{Discussion}
\label{sec:discussion}

The positive correlation between $q$ and A$_{\rm{wiggle}}$ is unambiguous and expected due to the physical origin of the wiggles. There is a correlation between the wiggles, in velocity space, and the spiral density waves, in physical space~\citep{longarini}. Given this, larger spirals in space will to some degree correspond with larger velocity perturbations. For example, more massive discs, with lower $m$-modes and therefore fewer spirals that dominate the morphology, are expected to have a larger wiggle amplitude. However, it is important to note that the amplitude of the wiggle is determined by how large a velocity shift the material in that spatial location has received, rather than the spatial scale of the spiral causing the wiggle.    

The positive correlation between $q$ and $\lambda_{\rm{wiggle}}$ is easy to understand. \citet{cossins2009} showed that $kH\simeq1$, where $k$ is the radial wavenumber and $H$ the height of the disk. The wavelength of the perturbation, i.e. the wavelength of the wiggle, is thus proportional to $H$ that, in a self-gravitating disc, scales as the disk to star mass ratio $q$ \citep{Kratter16}.

\par

\par
\begin{figure*}
    \centering
    \includegraphics[width=0.95\linewidth]{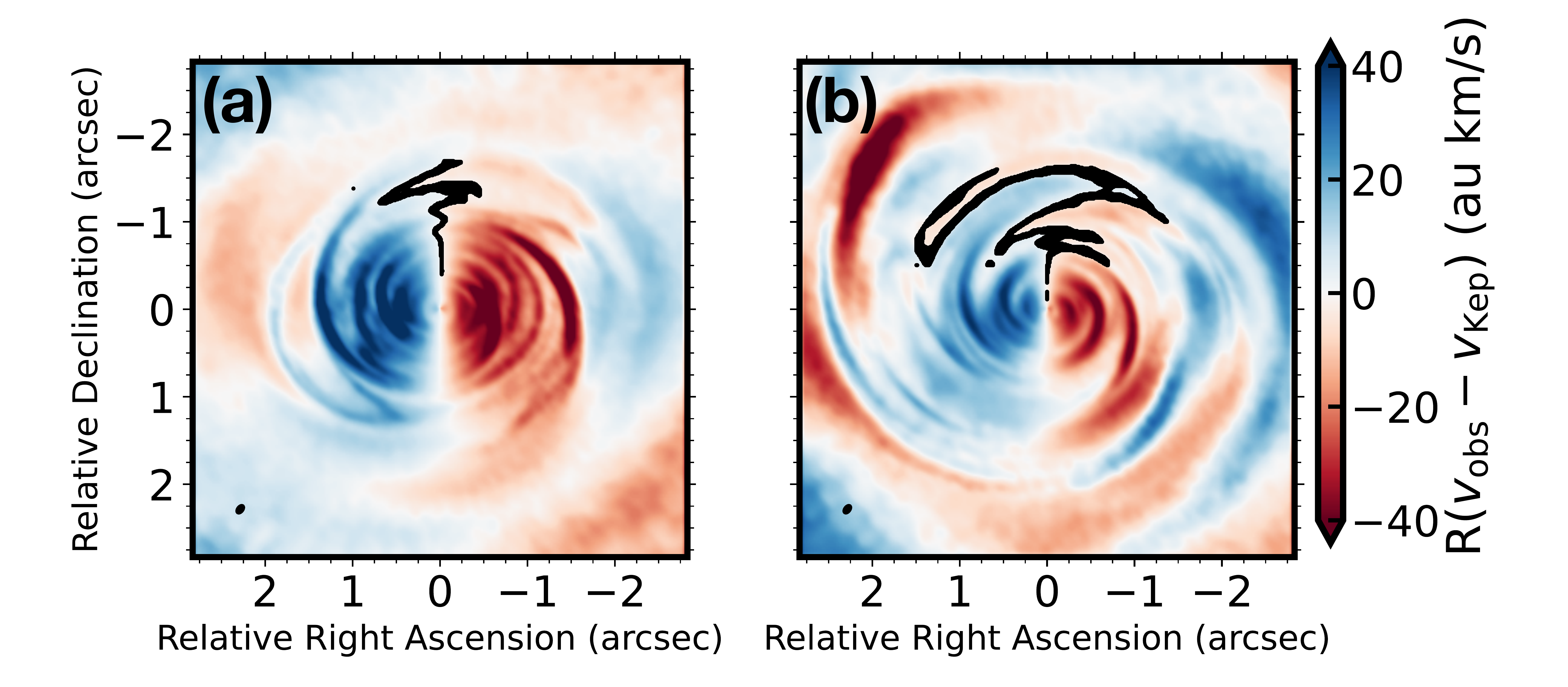}
    \caption{Moment-1 maps with Keplerian rotation subtracted and multiplied by the radius from the central sink, R (in au), for clarity. Using disc-to-star mass ratios of (a) $q=0.25$, (b) $q=0.75$. The ``interlocking fingers" that are predicted signatures of GI are present although clearly stronger for the more massive disc. The black line indicates where $v_{\rm{obs}} = v_{\rm{Kep}}$.}
    \label{fig:moments}
\end{figure*}

Figure~\ref{fig:moments} shows moment-1 maps with Keplerian rotation subtracted (with each pixel scaled by its distance from the central sink for clarity) for mass ratios of 0.25 and 0.75. These moment-1 maps shows the intensity-weighted velocity, which is calculated by Equation~\ref{eq:moment_1}

\begin{equation}
    \langle v \rangle = \frac{\int_{-\infty}^{\infty} v \mathrm{I}\left( v \right) \mathrm{d}v}{\int_{-\infty}^{\infty}\mathrm{I}\left( v \right) \mathrm{d}v}.
    \label{eq:moment_1}
\end{equation}

Both discs in Figure~\ref{fig:moments} exhibit the ``interlocking finger" pattern predicted by \citet{hall20} as a characteristic imprint of GI. However, the more massive disc, $q=0.75$, clearly has more pronounced perturbations, strongly suggesting that, when compared to the less massive disc, GI has a larger influence on its dynamics.

\subsection{Limitations and considerations}

Our study comes with a number of limitations. We vary only mass ratio and number of SPH particles between runs, exploring only the GI wiggle's dependence on disc-to-star mass ratio, and not on other parameters. This is likely a complicated issue, since, for example, discs are more stable around low-mass stars \citep{Haworth2020, Cadman2020}. Another issue is the cooling parameter, $\beta$, which we do not vary. Here, we have shown that for a constant $\beta$, $A_{\rm{wiggle}}$ depends on $q$. Our work, therefore, is essentially assuming that most discs undergoing GI will have similar radiative properties. Since $\delta\Sigma\propto\beta ^{-1/2}$, we expect that smaller $\beta$ gives stronger perturbations, thus wiggles with bigger amplitude.~\citet{longarini} provide an analytical model for the GI-wiggle, confirming this trend. Even had we varied $\beta$, this would not accurately reflect reality, since heating and cooling are complicated processes best considered by a full radiative transfer treatment. However, assuming constant $\beta$ allowed us to remove one variable from the analysis, which would not have been possible with full radiative transfer. 

We do not include the effect of viscosity on the damping of the GI wiggle in our analysis, but we can provide an order of magnitude estimate of its effect. In a non-viscous, non-irradiated disc, the amplitude of the density perturbation due to gravitational instabilities is such that the effecting heating provided by the instability balances radiative cooling \citep{Kratter16}. The effective heating associated to GI can be parameterized through an effective $\alpha_{\rm GI}$, proportional to the density perturbation:
\begin{equation}
    \alpha_{\rm GI}=\frac{4}{9\gamma(\gamma-1)}\frac{1}{\beta}\approx \frac{4}{9\gamma(\gamma-1)}\left(\frac{\Delta\Sigma}{\Sigma}\right)^2,
\end{equation}
where $\gamma$ is the ratio of specific heats and in the last step we have used the fact that for a non-viscous self-regulated disc $(\Delta\Sigma/\Sigma)^2\approx 1/\beta$ \citep{cossins2009}. If an additional viscosity is present, parameterized by a Shakura-Sunyaev coefficient $\alpha_{\rm SS}$, thermal balance is then rewritten as:
\begin{equation}
    \alpha_{\rm GI}+\alpha_{\rm SS} = \frac{4}{9\gamma(\gamma-1)}\left(\frac{\Delta\Sigma}{\Sigma} \right)^2 + \alpha_{\rm SS} = \frac{4}{9\gamma(\gamma-1)}\frac{1}{\beta},
\end{equation}
from which we obtain
\begin{equation}
   \left( \frac{\Delta\Sigma}{\Sigma}\right)^2 = \frac{1}{\beta}\left(1-\frac{9\gamma(\gamma-1)\alpha_{\rm SS}\beta}{4}\right).
    \label{eq:viscosity}
\end{equation}
Since the velocity perturbations are proportional to the density perturbations \citep{longarini}, we expect a similar reduction also in the amplitude of the wiggle, when $\alpha_{\rm ss}$ is non-negligible. Note that Equation (\ref{eq:viscosity}) is analogous to equation (15) in \citet{Rice11}, derived for irradiated discs, and in fact a similar reduction in the wiggle amplitude is also expected when one adds also irradiation as a source of disc heating.

\par
In this work, we intended to determine if it was possible to characterise the GI wiggle from current ALMA observations. We therefore included dust in our simulations since the one-fluid method of \citet{laibeprice2014a, laibeprice2014b, laibeprice2014c} only results in a very small slow-down ($\sim$ 5\%) of the code, so it is possible to perform a much more realistic, observationally motivated calculation at very little extra computational cost. GI discs trap dust in their spiral arms \citep{rice2004}, resulting in a varied radial dust distribution which affects the thermal disc structure in the radiative transfer calculation. This in turn can affect the intensity of molecular line emission, which is why we opted to use dust in our calculations. However, it would be interesting in future to vary both dust-to-gas ratio and dust grain distribution to explore what effect, if any, it has on the GI wiggle.

\par
An important point to note is that the resolution used in our analysis represents some of the highest spectral and spatial resolutions found in observations to date. Our convolved results had a spectral resolution of 0.03 km/s after Hanning-smoothing and were spatially convolved with a Gaussian beam of size $0.11\arcsec \times 0.07\arcsec$. While this resolution has been used on planet-containing discs~\citep{Pinte2018}, these objects are far below the mass threshold required for GI to be active. The current spatial and spectral resolution of observations of the best known GI candidate, Elias 2-27~\citep{Perez2016, paneque2021, veronesi2021}, is roughly a factor 3 below the resolution we use here. Further ALMA observations are therefore required to apply the method we describe here.

\section{Conclusions}
\label{sec:conclusion}

We used numerical simulations to determine a positive linear relationship between the amplitude of the GI Wiggle and disc-to-star mass ratio $q$, for a constant cooling parameter, $\beta$. The best fit relationship using $\beta=15$ is A$_{\rm{wiggle}}=50q-5.4$. The R$^{2}$ value from this fit is 0.88, suggesting that a linear fit is appropriate. A similar linear regression on the GI wavelength gave $\lambda_{\rm{wiggle}}=129q+66$ with an R$^{2}$ value of 0.89. This also indicates that approximating the GI wavelength as a linear function of the mass ratio is also valid. We present a heuristic argument based both on previous findings and physical reasoning to support our numerical results.

\par
Our results hold for convolution to a spectral resolution of 0.033 km/s,  and spatial convolution using a Gaussian beam of size $0.11\arcsec \times 0.07\arcsec$. This indicates that determination of wiggle wavelength and amplitude from ALMA observations with maximum resolution is immediately possible. We therefore suggest that our results can be used to constrain disc mass in systems that contain the GI Wiggle.

\section*{Acknowledgements}
We thank the referee for their review, which greatly improved the clarity of the paper. We thank Cathie Clarke and Sahl Rowther for discussions that improved the accuracy of this paper. This study was supported in part by resources and technical expertise from the Georgia Advanced Computing Resource Center, a partnership between the University of Georgia’s Office of the Vice President for Research and Office of the Vice President for Information Technology. Funding for MCFOST was provided by the Australian Research Council under contracts FT170100040 and DP180104235 and from Agence Nationale pour la Recherche (ANR) of France under contract ANR-16-CE31-0013. This project has received funding from the European Union's Horizon 2020 research and innovation programme under the Marie Sklodowska-Curie grant agreement No 823823 (Dustbusters RISE Project). BV acknowledges funding from the ERC CoG project PODCAST No 864965.

\section*{Data availability}
The data used in this article will be shared on reasonable request to the corresponding author on a collaborative basis of coauthorship.


\bibliographystyle{mnras}
\bibliography{bib} 



\section*{Appendix} \label{sec:app}
\appendix

\begin{figure}
 \includegraphics[width=\linewidth]{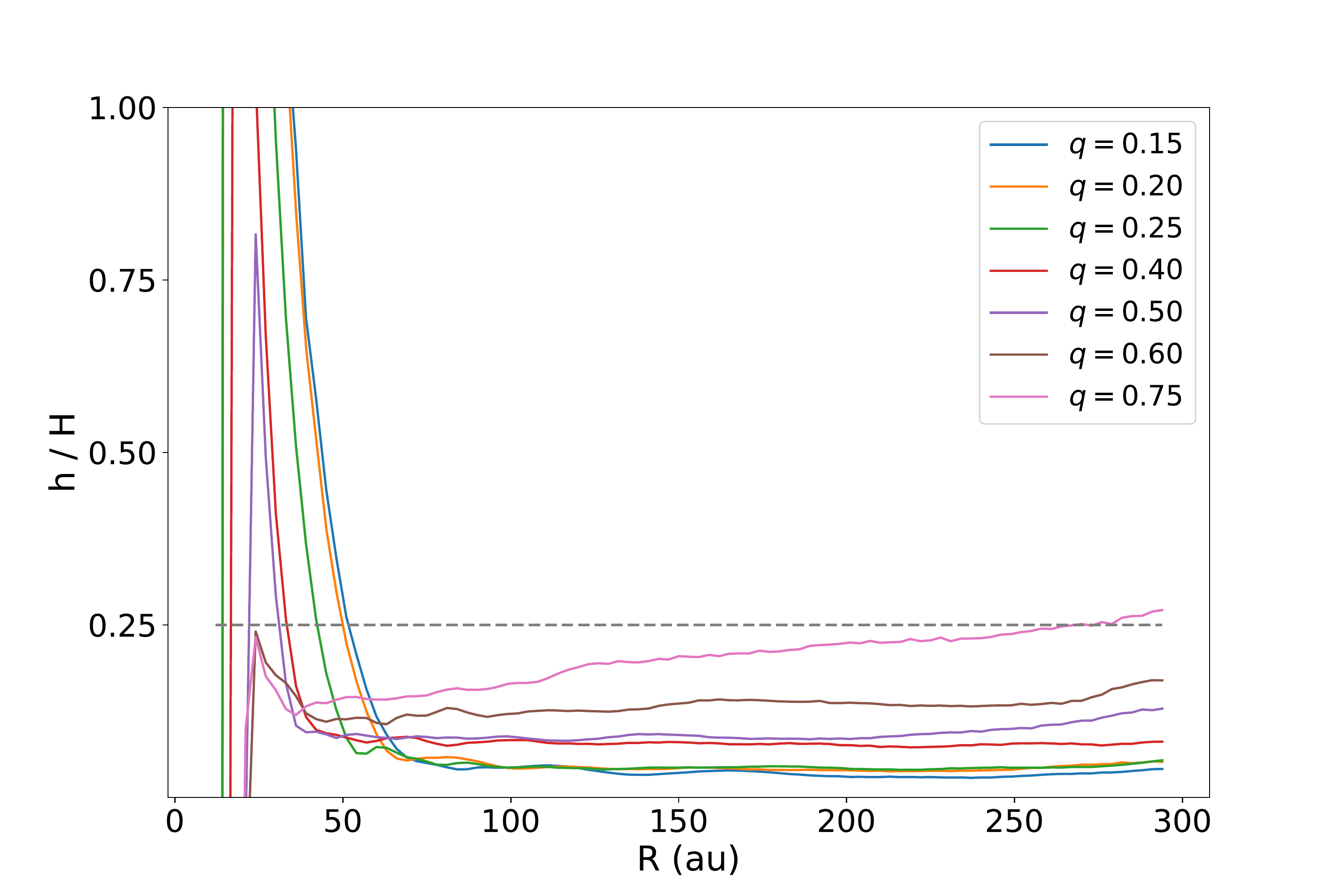}
 \caption{h/H for all $q$ within R$\leq300$ au, which is greater than all $R_{\rm{outer}}$. N=5e5, 5e5, 7.5e5, 1e6, 1.25e6, 1.25e6, 1.5e6.}
 \label{fig:h_H}
\end{figure}

A wide range of disc masses was simulated using varying SPH particle number. We ensured that all simulations were resolved according to the criterion presented in \citet{Nelson2006}, such that at least 4 SPH particle smoothing lengths per scale height (i.e. h/H $\leq 0.25$) within the considered radii are maintained (Appendix Figure~\ref{fig:h_H}).

\begin{figure}
 \includegraphics[width=0.9\linewidth]{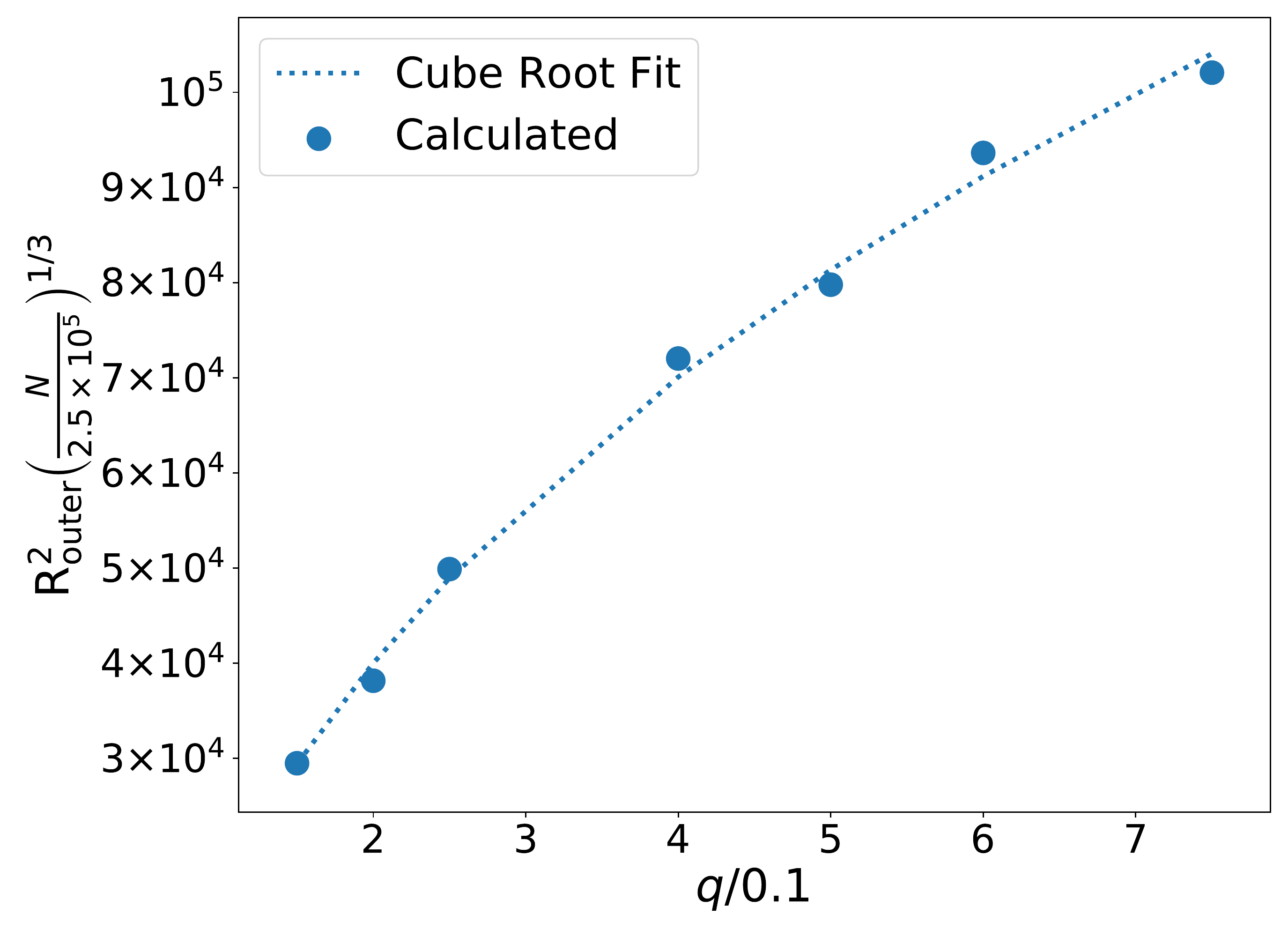}
 \caption{Verifying the relationship that gives a measure of the consistency of artificial viscosity between mass ratios. N=5e5, 5e5, 7.5e5, 1e6, 1.25e6, 1.25e6, 1.5e6 (Table~\ref{tab:r_outer}).}
 \label{fig:cube_root_fit}
\end{figure}

We attempt to use a number of SPH particles, N, for each simulation such that the artificial viscosity is approximately constant between mass ratios. \citet{lodatoclarke2011} found that $$\frac{h}{H} = \frac{5\times 10^{-3}}{m(R)}\left(\frac{q}{0.1}\right)^{1/3}\left(\frac{\rm{N}}{2.5 \times 10^{5}}\right)^{-1/3}$$ where $m(R) = \Sigma R^{2}/\rm{M}_{*}$. Given that $h/H \propto \alpha_{\rm{visc}}$, M$_{\*}$ is constant for all simulations, and, at $R_{\rm{outer}}$, $\Sigma$ is small in relation to the interior of the disc, a cube root relationship between $\rm{R}_{\rm{outer}}^{2}\left(\frac{N}{2.5\times10^{5}}\right)^{1/3}$ and $q/0.1$ would indicate that we have at least roughly achieved this objective. As Appendix Figure~\ref{fig:cube_root_fit} shows, a cube root fit is approximately correct.

\vspace{4mm}

\bsp	
\label{lastpage}
\end{document}